\documentclass{ws-procs9x6}

\begin{document}

\title{Meet, Discuss and Trust each other: large versus small groups}

\author{Timoteo Carletti$^1$}

\address{D\'epartement de Math\'ematique, Facult\'es Universitaires Notre Dame
  de la Paix\\ Namur, B5000, Belgium\\
$^1$E-mail: timoteo.carletti@fundp.ac.be}

\author{Duccio Fanelli$^{2}$}

\address{Dipartimento di Energetica and CSDC, Universit\`a di Firenze, and
  INFN,\\ Firenze, 50139, Italy\\
$^{2}$E-mail: duccio.fanelli@gmail.com}

\author{Alessio Guarino$^{3}$}

\address{Universit\'e de la Polyn\'esie Francaise\\BP 6570 Faa'a, 98702, French
Polynesia\\
$^{3}$E-mail: alessio.guarino@upf.pf}

\author{Andrea Guazzini$^{4}$}

\address{Institute for Informatics and Telematics (IIT)\\ Pisa, 56124,Italy\\
$^{4}$E-mail: andrea.guazzini@unifi.it}

\begin{abstract}
In this paper we propose a dynamical interpretation of the sociological
distinction between large and small groups of interacting individuals. In 
the former case individual behaviors
are largely dominated by the group effect, while in the latter mutual
relationships do matter. Numerical and analytical tools are combined to
substantiate our claims.
\end{abstract}


\bodymatter

\section{Introduction}

Sociophysics is a long standing~\cite{GGS} research field addressing
issues related to the characterization of 
the collective social behavior of individuals, such as culture
dissemination, the spreading of linguistic conventions, and
the dynamics of opinion formation~\cite{BetAl,Deffuant,PetAl,Sznajd,SS}. These are all interdisciplinary applications which sit the interface of different domains. The challenge is
in fact  
to model the dynamical evolution of an ensemble made of interacting,
micro--constituents and infer the emergence of collective, macroscopic
behaviors that are then 
eventually accessible for direct experimental inspection.
Agent based computational models are widely employed in sociophysics applications and also adopted in this paper. They provide in fact a suitable setting to define local
rules which govern the evolution of the microscopic constituents.\\
In recent years, much effort has been devoted to the investigation of social
networks, emerging from interaction among humans. In the
sociological literature a main distinction has been drawn between
small~\cite{bion} and large~\cite{Bon,McDougall,Berk} groups, as depending on its intrinsic size, the system apparently exhibits distinct social behaviors. Up to a number of 
participants of the order of a dozen, a group is considered small. All members have a
clear perception of other participants, regarded as individual entities: Information hence flows because of mutual relationships. Above this reference
threshold, selected individuals see the vast majority of the group as part of a uniform mass: There is no perception of the individual 
personality, and only average behaviors matter. This
distinction is motivated by the fact that usually humans act
following certain prebuilt \lq\lq schemes\rq\rq~\cite{FiskeNeuberg,Neuberg} resulting from past experiences, which enables for rapid 
decision making whitout having to necessarily screen all available data. This innate data analysis process 
allows one for a dramatic saving of cognitive resources.\\
These conclusions have been reached on the basis of empirical and
qualitative evidences~\cite{bion,Bavelas1,Bavelas2,Leavitt}. We are here
interested in detecting the emergence of similar phenomena in a simple model of opinion dynamics~\cite{pre,epjb}. 
As we shall see in our proposed formulation agents posses a continuous opinion on a given subject and possibly modify 
their beliefs as a consequence of binary encounters.\\
The paper is organized as follows. In the next section the model is
introduced. Forthcoming sections are devoted to characterizing the quantities being inspected and develop the mathematical
 treatment. In  the final section we sum up and comment about future perspectives. 

\section{The model}

\begin{figure}[htbp]
\centering
\includegraphics[scale=0.3]{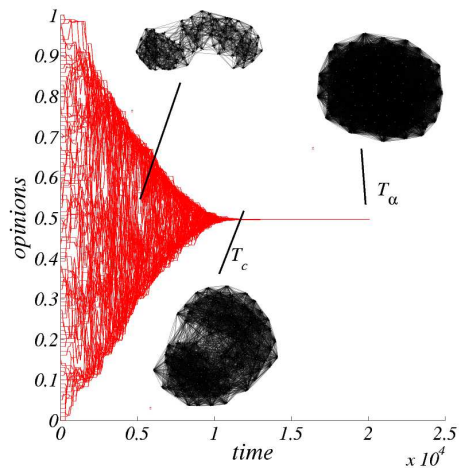}
\caption{Opinions as function of time. The underlying network is displayed at 
  different times, testifying on the natural tendency to evolve towards
  a coherent ensemble of affine individuals. $T_c$ and 
  $T_{\alpha}$ are measured according to our conventions.}
\label{fig:fig1}
\end{figure}

We shall investigate the aforementioned effects related to the size of the
group of interacting individuals ({\it social group}), within a simple opinion
dynamics model, recently introduced in~\cite{pre} . For the sake of 
completeness we hereafter recall the main ingredients characterizing the model. The interested 
reader can refer to~\cite{pre} for a more detailed account.\\
We consider a closed group composed by $N$ individuals, whose opinion on a
given issue is represented by a continuous variable $O_i$, scanning the
interval $[0,1]$; moreover each agent $i$ is also characterized by the
so--called {\em affinity}, 
a real valued vector of $N-1$ elements, labeled $\alpha_{ij}$, which measures the quality of the
relationship between $i$ and any other actor $j$ belonging to the community. \\
Agents interact via binary encounters possibly updating their opinion and
relative affinity, which thus evolve in time. Once agents $i$ and $j$ interact,
via the mechanism described below, they converge to the mean opinion value, 
provided their mutual affinity scores falls below a pre-assigned threshold quantified via the 
parameter $\alpha_c$. In formulae:
\begin{equation}
O_i^{t+1} = O_i^{t}- \frac{1}{2} \, \Delta
O_{ij}^{t}\,\Gamma_1\left(\alpha^t_{ij}\right)  \quad \& \quad O_j^{t+1} =
O_j^{t}- \frac{1}{2} \, \Delta O_{ji}^{t}\, \Gamma_1\left(\alpha^t_{ji}\right)
\, ,  
\label{eq:oevol}
\end{equation}
where $\Delta O^t_{ij}=O_{i}^t-O_{j}^t$ and $\Gamma_1 \left(x\right)=
\frac{1}{2}\left[ \tanh (\beta_1 (x-\alpha_c)) + 1 \right]$. The latter works as an {\em
  activating function} defining the region of trust for effective social interactions.
  On the other hand bearing close enough opinions on a selected topic, might
  induce an enhancement of the  
  relative affinity, an effect which is here modeled as: 
  \begin{equation}
\alpha_{ij}^{t+1} = \alpha_{ij}^{t} + \alpha_{ij}^{t}
      (1-\alpha_{ij}^{t}) \,\Gamma_2 \left(\Delta O^t_{ij}\right) \quad \& \quad
\alpha_{ji}^{t+1} = \alpha_{ji}^{t} + \alpha_{ji}^{t}
      (1-\alpha_{ji}^{t}) \,\Gamma_2 \left(\Delta O^t_{ji}\right) \, ,
\label{eq:alphaevolv}
\end{equation}
being $\Gamma_2 \left(x\right)= -\tanh(\beta_2 (|x| - \Delta O_c))$. This
additional  {\em activating function} quantifies in $\Delta O_c$ the largest
difference in opinion  
($\Delta O^t_{ij}$) which yields to a positive increase of the affinity amount $\alpha^t_{ij}$.  
The parameters $\beta_1$ and $\beta_2$ are chosen large 
enough so that $\Gamma_1$ and $\Gamma_2$ are virtually behaving as
step functions. Within this 
working assumption, the function $\Gamma_1$ assumes value $0$ or $1$, while
$\Gamma_2$ is alternatively $-1$ or $+1$, depending on the value of their
arguments~\footnote{We shall also emphasize that the logistic contribution
  entering~\eref{eq:alphaevolv} maximizes the change in affinity when
  $\alpha_{ij}^t \approx 0.5$, corresponding to agents $i$ which have not yet
  build a definite judgment on the selected interlocutor $j$. Conversely, when
  the affinity is close to the boundaries of the allowed domain, marking a
  clear view on the worth of the interlocutor, the value of  $\alpha_{ij}^t$
  is more resistant to subsequent adjustments.}.\\
The affinity variable, $\alpha_{ij}^t$, schematically accounts
for a large number of hidden traits (diversity,personality, attitudes,
beliefs...), 
which are nevertheless non trivially integrated as an abstract simplified form
into the 
model. Note also that the affinity accounts for a {\em memory}
mechanism: indeed once two agents 
meet, the outcome of the interaction in part depends on
their history via the affinity scores.  \\
To complete the description of the model let us review the 
selection rule here implemented. Each time step a first agent $i$, is randomly
extracted, with uniform probability.  Then a second agent
$j$ is selected, which   minimizes the {\em social metric} $D_{ij}^t$ and time
$t$. This is a quantity defined as:
\begin{equation}
  \label{eq:socmetr}
  D_{ij}^t = d^t_{ij}+\mathcal{N}_j(0,\sigma)\, ,
\end{equation}
where $d^t_{ij}=|\Delta O_{ij}^t|
(1-\alpha_{ij}^t)$ is the so--called {\em social distance}  and 
$\mathcal{N}_j(0,\sigma)$ represents 
a normally distributed noise with zero mean and
variance $\sigma$, that can be termed
{\em social temperature}~\cite{pre}. The rationale inspiring the mechanisms
here postulated goes as follows:
The natural tendency for agent $i$ to pair with her/his closest homologous
belonging to the community (higher affinity, smaller opinion distance), is 
perturbed by a stochastic disturbance, which is intrinsic to the social environment 
(degree of mixing of the population).\\
The model exhibits an highly non linear dependence on the involved parameters,
$\alpha_c$, $\Delta O_c$ and $\sigma$. In a previous work~\cite{pre} the
asymptotic behavior of the opinions dynamics was studied and the existence 
of a phase transition between a consensus
state and a polarized one demonstrated. It should be remarked however that
the fragmented case might be metastable; in fact if the mean separation between the adjacent opinion peaks
is smaller than the opinion interaction threshold, $\Delta O_c$, there always
exists a finite, though small, probability of selecting two individuals
belonging to different clusters, hence producing a gradual increase in the
mutual affinities, which eventually lead to a merging of the, previously,
separated clusters. This final state will be achieved on extremely long time
scales, diverging with the group size: socially relevant dynamics are hence
likely to 
correspond to the metastable regimes.\\
A typical run for $N=100$ agents is reported in
the main panel of Fig.~\ref{fig:fig1}, for a choice of the parameters which
yields to a monoclustered phase. This is the setting that we shall be focusing
on in the forthcoming discussion: Initial opinions are uniformly distributed
in the interval $[0, 1]$, while $\alpha_{ij}^0$ are randomly assigned in
$[0,1/2]$ with uniform distribution, parameters have been fixed to
$\alpha_c=0.5$, $\Delta O_c=0.5$ and $\sigma = 0.01$.\\
Once the cluster is formed, one can define the {\em opinion convergence time},
$T_c$, i.e. time needed to aggregate all the agents to the main opinion
group~\footnote{We assume that a group is formed, i.e. aggregated, once the
  largest difference between opinions of agents inside the group is smaller
  than a threshold, here $10^{-4}$.}. A second quantity $T_\alpha$ can be
introduced,  
which measures the time scale for the convergence of the {\em mean group
  affinity} to its asymptotic value $1$.  
The latter will be rigorously established in the next section.  \\
Such quantities are monitored as function of time and results are
schematically reported in Fig.~\ref{fig:fig1}. As 
clearly depicted, the evolution occurs on sensibly different time scales, 
the opinion converging much faster for the set of parameters here employed. \\
In the remaining part of this paper, we will be concerned with analyzing 
the detail of this phenomenon highlighting the existence of different regimes 
as function of the amount of simulated individuals. More specifically,
we shall argue that in small groups, the mean
affinity converges faster than opinions, while the opposite holds in a 
large community setting. Our findings are to be benchmarked with the empirical evidences, 
as reported in the psychological literature. 
It is in fact widely recognized that the dynamics of a small group (workgroup)   
proceeds in a  two stages fashion: First one learns about colleagues to evaluate 
their trustability, and only subsequently weight their input 
to form the basis for decision making. At variance, in large
communities, only a few binary interactions are possible among selected participants
within a reasonable time span. It is hence highly inefficient to 
wait accumulating the large number of information that would 
eventually enable to assess the reliability of the interlocutors.
The optimal strategy in this latter case is necessarily (and solely) bound to 
estimating the difference in opinion, on the topic being debated.   
\section{The social network of affinities}
\begin{figure}[htbp]
\begin{center}
\mbox{\includegraphics[scale=0.25]{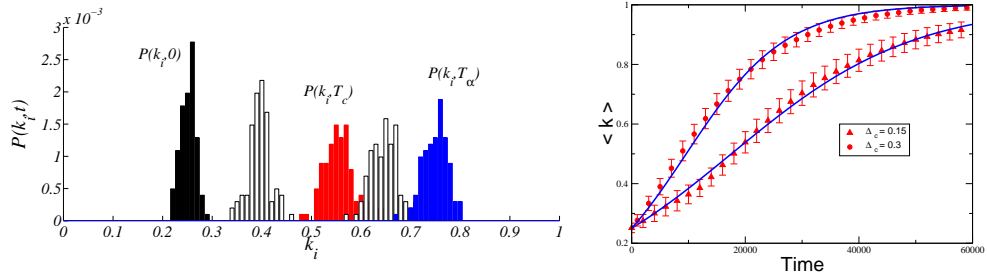}\quad
\includegraphics[scale=0.21]{evol_k.eps}}
\end{center}
\caption{Time evolution of the degree probability distribution
  function and the averaged degree. Left panel : Several  
  histograms representing successive snapshots of the dynamics are displayed: 
  $t=0$ (black online), two generic intermediate
  times (wight),  $t=T_c$ (red online) and $t=T_{\alpha}$ (blue
  online); histograms are normalized to unity. Right panel : $<k>$ versus
  time. Symbols refer to direct simulations. The solid lines are obtained as a
  one parameter fit of the theoretical expression~\eref{eq:eq3}.}   
\label{fig:fig2}
\end{figure}

In our model the affinity enters the selection mechanism that
makes agents to interact. We can restate this fact by postulating the
existence of an underlying {\em social network}, which  
drives the interactions
and thus the opinion flow. In this perspective, the
affinity can be seen as the {\em
  adjacency matrix} of a {\em weighted}~\footnote{In fact the trustability
  relation is measured in terms of the \lq\lq weights\rq\rq $\alpha_{ij}^t\in
  [0,1]$.} graph. In such a way we are formally dealing with an {\em adaptive} social 
  network~\cite{GrossBlasius,ZESM} : The network topology
influences the opinion dynamics, the latter providing a feedback on the 
network itself. In other words, the evolution of the topology is inherent to the dynamics of the model because 
of the  proposed self-consistent formulation and 
not imposed a priori as an additional, external ingredient, i.e. rewire and/or
add/remove links according to a given probability~\cite{HN,KB} once the
state variables have been updated.\\
From this point of view, the mean group affinity is the {\em
  averaged outgoing degree} --  called for short "the degree" in the following
-- of the network:
\begin{equation}
  \label{eq:degree}
<k>(t)=\frac{1}{N}\sum_i k_i^t\, ,
\end{equation}
where the degree of the i--th node is
$k_i^{t}=\sum_{j}\alpha_{ij}^t/(N-1)$. The normalizing factor $N-1$ allows for
a direct  
comparison of networks made of a different number of agents. Let us observe
that we chose to normalize with respect to $N-1$ because no self-interaction
is allowed for.  \\ 
In the left panel of Fig.~\ref{fig:fig2} we report the
probability distribution function for the degree, as a function of
time. Let us observe that the initial distribution is correctly centered
around the value $1/4$, due to the specificity of the chosen initial
condition. The approximate Gaussian shape results from a straightforward
application of the Central Limit Theorem to the variables
$(k_i^t)_{i=1,\dots,N}$. In the right panel of Fig.~\ref{fig:fig2} the time
evolution of the mean 
degree $<k>(t)$ is reported.  Starting from the initial value $1/4$, the mean
degree increases and eventually reaches the value $1$, characteristic of a 
complete graph. As previously mentioned this corresponds to a social network 
where agents are (highly) affine to each other.  In the same panel we
also plot the 
analytical curve for $<k>(t)$, as it is determined hereafter.\\

From the previous observation, it is clear that the time of equilibration of
$<k>(t)$  provides an indirect measure of the convergence time for $\alpha_{ij}^t\rightarrow
1$. The affinity convergence time, $T_{\alpha}$ can be thus defined as: 
\begin{equation} 
\label{eq4}
T_{\alpha}=\min \{t>0 :  <k>(t) \geq \eta\} \, ,
\end{equation} 
where $\eta\in (0,1)$ is a threshold quantity. The closer $\eta$ is to $1$ the 
stronger the condition on the value of $\alpha_{ij}^t$ (the larger the value for $T_{\alpha}$) which identifies an affine 
unit. In the following we shall assume $\eta = 3/4$.
\section{Results}

\begin{figure}[htbp]
\centering
\includegraphics[scale=0.31]{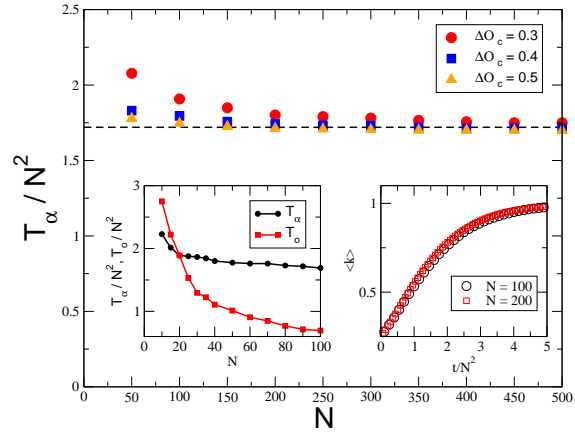}
\caption{Main panel: $T\alpha/N^2$ vs $N$ for different values of the
  parameter $\Delta O_c$. The data approach a constant value ($T_{\alpha} /N^2
  \sim 1.72$) clearly indicating that the time of convergence of the affinity
  matrix scales quadratically with the number of agents, in agreement with the
  theory. The asymptotic value estimated by our theory is $2.19$, the
  discrepancy being therefore quantified in about $15$~\%. Left inset:
  $T_\alpha/N^2$ and $T_c/N^2$ vs $N$ for $\Delta O_c=0.5$. As predicted by 
the theory and the numerics a crossover is found for
groups for which opinions converge slower than the affinities: this is 
the signature of a distinctive difference in the behavior of small and large
groups, numerically we found that this difference is effective for $N\sim
20$. Right inset: $<k>$ vs $t/N^2$ is plotted for two different values of
$N$. As expected the two curves nicely collapse together.}
\label{fig:fig_conv}
\end{figure}

The aim of this section is to analyze the behavior of $T_{\alpha}$ and of
$<k>(t)$ so to provide an analytical interpretation for the results presented
in the previous section.\\
From the definition of mean degree~\eref{eq:degree} one can compute the time
evolution of 
$<k>(t)$ in a continuous time setting:
\begin{equation} 
\label{eq:tevol}
\frac{d <k>}{d t} = \frac{1}{N (N-1)} \sum_{i,j} \frac{d\alpha^t_{ij}}{d t}\, . 
\end{equation} 
To deal with the update rule for the affinity~\eref{eq:alphaevolv}
we assume that we can decouple the opinions and affinity dynamics 
by formally replacing the activating function $\Gamma_2(\Delta O_{ij}^t)$ with a suitably tuned
constant. Strictly speaking, this assumption is correct only for $\Delta O_c=1$,
in which case the opinions are always close enough so to magnify the mutual
affinity scores of the interacting agents, as a results of a self consistent
evolution. For the general case, $\Delta O_c \leq 1$, what
we are here proposing is to replace $\Gamma_2(\Delta O_{ij}^t)$ by some {\em
  effective value} termed $\gamma_{ef\!f}$, determined by the dynamics itself. The technical development that yields to the 
  optimal estimate of $\gamma_{ef\!f}$ will be dealt in the appendix.\\
Under this assumption, the  equation~\eref{eq:alphaevolv} for the evolution of $\alpha^t_{ij}$ admits the following continuous version :
\begin{equation}
\label{eq:alphaevol}
  \frac{d\alpha^t_{ij}}{d t}=\gamma_{ef\!f} \alpha^t_{ij}
  \left(1-\alpha^t_{ij}\right)\, ,
\end{equation}
which combined to equation ~\eref{eq:tevol} returns the following equation for the mean degree evolution :
\begin{equation} 
\label{eq1}
\frac{d <k>}{d t} = \gamma_{ef\!f} \left( <k>(t) -
  <(\alpha_{ij}^t)^2>\right)\, .  
\end{equation} 
Assuming the standard deviation of
$(\alpha_{ij}^t)$ to be small~\footnote{This assumption is supported by
  numerical simulations not reported here and by the analytical considerations
presented in~\cite{teo}.} for all $t$, implies $<~(\alpha_{ij}^t~)^2> \sim <\alpha^t_{ij}>^2= <k>^2$, which allows us to cast the 
previous equation for $<k>$ in the closed form :
\begin{equation} 
\label{eq2}
\frac{d <k>}{d t} = \gamma_{ef\!f} \left( <k> - <k>^2 \right)\, . 
\end{equation}
This equation can be straightforwardly solved to give:
\begin{equation} 
\label{eq:eq3}
<k> = \frac{k_0}{k_0+(1-k_0) e^{-\gamma_{ef\!f} t}}\, ,
\end{equation} 
where $k_0=<k>(0)$. We can observe that such solution provides the correct
asymptotic value for large $t$. Moreover $\gamma_{ef\!f}$ plays the role
of a {\em characteristic} time and in turn enables us to quantify the convergence time $T_\alpha $
of the affinity via :
\begin{equation} 
\label{eq5}
T_\alpha = \frac{1}{\gamma_{ef\!f}} \log \left( \frac{\eta (1-k_0)
  }{k_0(1-\eta)} \right)=\frac{\eta^{\prime}}{\gamma_{ef\!f}} \, .
\end{equation} 
In the appendix we determine~\footnote{This is case a)
  of~\eref{eq:gammaefffinal}. In the second case the result is
  straightforward: $T_{\alpha}=\eta^{\prime}N^2/(\rho+1)$.} the following
relation which allows to 
express $\gamma_{ef\!f}$ as a function of the relevant variables and parameters
of the models, i.e. $T_c$, $\Delta O_c$ and $N$:
\begin{equation}
  \label{eq:gammed2}
  \gamma_{ef\!f}=\frac{1}{N^2}+\frac{T_c}{T_\alpha N^2} \rho\, ,
\end{equation} 
where $\rho=-(1+2 \Delta O_c \log(\Delta O_c) - \Delta O_c^2 )$, thus
recalling~\eref{eq5} we can finally get:
\begin{equation}
  \label{eq:Talpha}
 T_{\alpha} =  \eta^{\prime}N^2\left(1-
   \frac{T_c\rho}{N^2\eta^{\prime}}\right)\, . 
\end{equation} 
From previous works~\cite{propaganda,pre} we know that the dependence of $T_c$
on $N$, for large $N$, is less than quadratic. Hence, for large $N$, the second term in the parenthesis 
drops, and we hence conclude that, the affinity convergence time grows like
$T_{\alpha}\sim N^2$, as clearly shown in the main panel of
Fig.~\ref{fig:fig_conv}. The prefactor'e estimate is also approximately correct, as discussed in the caption of 
Fig.~\ref{fig:fig_conv}.\\
 The above results inspire a series of intriguing observation. First, it is implied that the larger the group size the bigger $T_{\alpha}$ with
respect to $T_c$. On the contrary, making $N$ smaller the gap progressively fades off. Dedicated numerical simulations (see left inset of
Fig.~\ref{fig:fig_conv}) allows to indentify a turning point which is reached
for small enough  values of $N$: there  
the behavior is indeed opposite, and, interestingly, $T_c > T_{\alpha}$. The transition here reproduced 
could relate to the intrinsic peculiarities of the 
so called \lq\lq small group dynamics\rq\rq to which we made reference in the
introductory section~\cite{bion,Bon,McDougall,Berk} . Furthermore, 
it should be stressed that the critical group size determining the switching between the two regimes here identified, 
is quantified in $N \simeq 20$, a value which is surprisingly closed to the one being quoted in social studies.

\section{Conclusion}

In this paper we study a model of continuous opinions dynamics already
proposed in~\cite{pre} , which incorporates as main ingredient the affinity
between agents both acting on the selection rule for the binary interactions
as well entering the postulated mechanism for the update of the individual
opinions.\\ 
Analyzing the model in the framework of adaptive networks we have been able to
show that the sociological distinction between large and small groups can be
seen as dynamical effect which spontaneously arises in our system. We have in
fact proven that for a set of realistic parameters there exists a critical
group size, which is surprisingly similar to the one reported in the
psychological literature. Below this critical value agents  
first converge in mutual affinity and only subsequently achieved a final
consensus on the debated issue. At variance, in large groups the opposite
holds~: The
convergence in opinion is the driving force for the aggregation process,
affinity converging on larger time scales. 

\appendix{Computation of $\gamma_{ef\!f}$}
\label{sec:calcGamma}

\begin{figure}[htbp]
\centering
\includegraphics[scale=0.23]{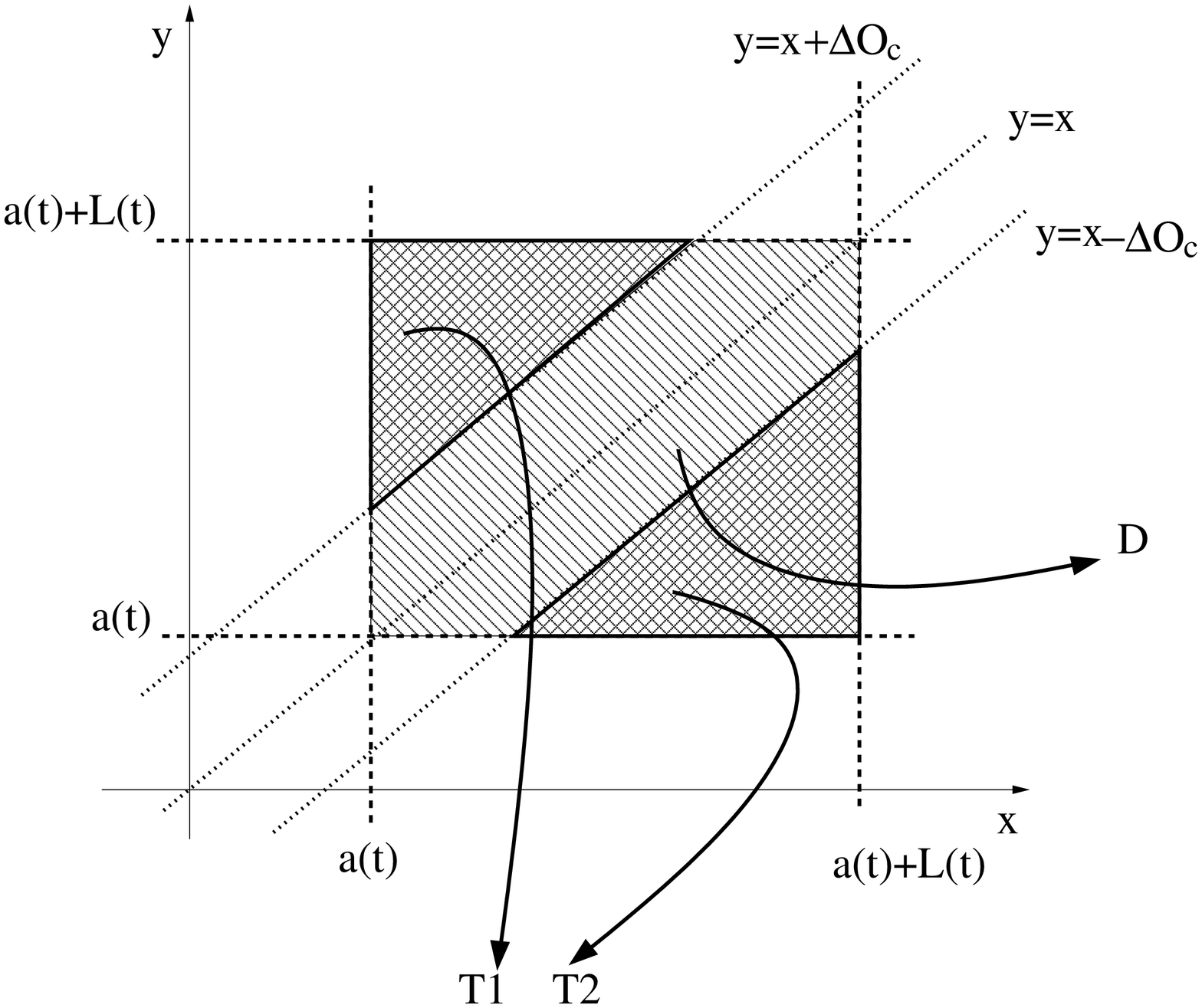}\hfill
\includegraphics[scale=0.23]{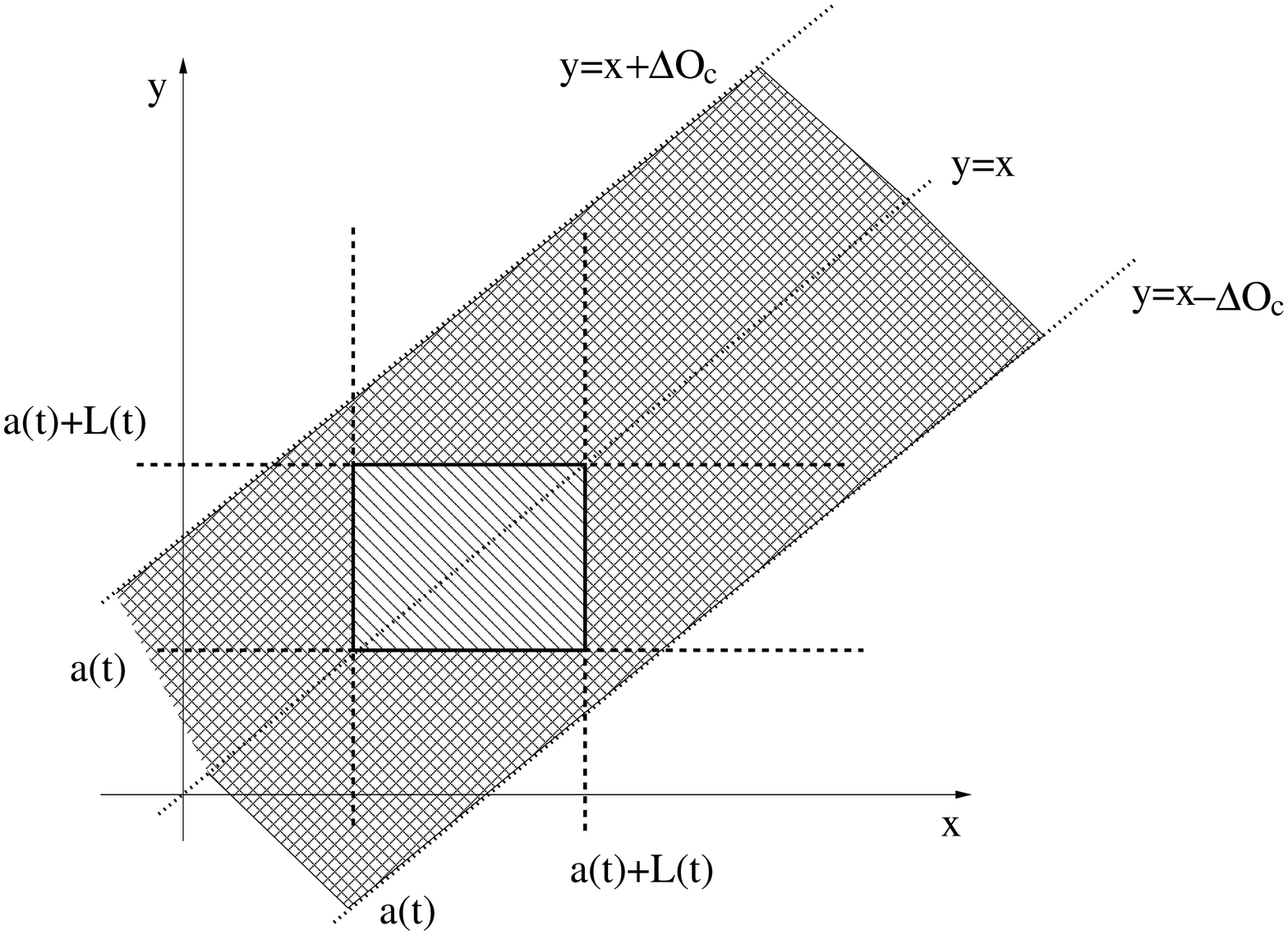} 
\caption{The geometry of the integration domains. On the left panel the case
  $L>\Delta O_c$, while on the right one the case $L< \Delta O_c$.}
\label{fig:chi}
\end{figure}

The aim of this paragraph is to provide the necessary steps to decouple the
opinion and affinity dynamics, by computing an effective value of the
activating function $\Gamma_2(\Delta O_{ij}^t)$, hereby called
$\gamma_{ef\!f}$. This will be obtained by first averaging $\Gamma_2$ with
respect to the opinions and then taking the time average of the resulting
function:
\begin{equation}
  \label{eq:gammaeff}
  \gamma_{ef\!f}=\lim_{t\rightarrow \infty}\frac{1}{t} \int_0^t d\tau\, 
  <\Gamma>_{op}(\tau) \, , 
\end{equation}
where the opinion--average is defined by: $$<\Gamma>_{op}(t)=\int_0^1 \, dx
\int_0^1 dy \, \Gamma_2(|x-y|) f(x)f(y),$$ being $f(\cdot)$ the opinions
probability distribution function. To perform the computation we will assume
that for each $t$, the opinions are 
uniformly distributed in the interval $[a(t),a(t)+L(t)]$ where $L(t)=1-t/T_c$
and $T_c$ is the opinion convergence time, hence $f(\cdot)=1/(NL(t))$. This
assumption is 
motivated by the \lq\lq  
triangle--like\rq\rq convergence pattern as clearly dsplayed in the main panel
of Fig.~\ref{fig:fig1}.\\
Assuming $\beta_2$ large enough, we can replace $\Gamma_2$ by a step
function. Hence: 
\begin{equation}
  \label{eq:gammaop2}
  <\Gamma>_{op}(t)=\frac{1}{N^2L^2(t)}\int_{a(t)}^{a(t)+L(t)} \, dx
  \int_{a(t)}^{a(t)+L(t)} \, dy \, \chi(x,y)\, , 
\end{equation}
where $\chi(x,y)$ is defined by (see also Fig.~\ref{fig:chi})
\begin{equation}
  \label{eq:defchi}
  \chi(x,y)=
  \begin{cases}
    1 & \text{if $|x-y|\geq \Delta O_c$, i.e. in the triangles
    $T_1\cup T_2=Q\setminus D$}\\
    -1 & \text{ otherwise, i.e. in $D$} \, ,
  \end{cases}
\end{equation}
where $Q$ is the square $[a,a+L]\times [a,a+L]$.\\
Let us observe that this applies only when $L(t)> \Delta O_c$ (see left panel
of Fig.~\ref{fig:chi}); while if $L(t)< \Delta O_c$ the whole integration
domain, $[a,a+L]\times [a,a+L]$, is contained into the $|x-y|<\Delta O_c$ (see
right panel of Fig.~\ref{fig:chi}). In this latter case, the integration turns out to be simpler. 
In other words the integral in~\eref{eq:gammaop2} corresponds to measure the
area, $|D|$, of the domains shown in Fig.~\ref{fig:chi} with a sign. Let us
perform 
this 
computation according to the two cases: $L>\Delta O_c$ or $L<\Delta O_c$.\\
In the case $L>\Delta O_c$, the area of $D$ is given by
$|D|=|Q|-|T_1|-|T_2|$, hence
\begin{eqnarray}
\label{eq:gamma1}
  <\Gamma>_{op}(t) &=&
-\frac{1}{N^2 L^2}\left(-|D|+|T_1|+|T_2|\right) = -\frac{1}{N^2 L^2}\left(-|Q|+2|T_1|+2|T_2|\right)
\\ 
&=&-\frac{1}{N^2 L^2}\left(-L^2+4\frac{(L-\Delta O_c)^2}{2}\right)= \frac{1}{N^2}\left[ 1-2\left(1-\frac{\Delta
      O_c}{L}\right)^2 \right] \quad\quad \text{(if $L>\Delta O_c$)}\notag\, . 
\end{eqnarray}
On the other hand if $L<\Delta O_c$, because the
square $Q$ is completely contained into the domain $|x-y|<\Delta O_c$ where
$\chi$ is equal to $-1$, we easily get: $<\Gamma>_{op}(t) = -\frac{1}{N ^2
  L^2}(-L^2)=\frac{1}{N^2}$, if $L<\Delta O_c$. This last
relation together with~\eref{eq:gamma1}, can be casted in
a single formula: 
\begin{equation}
  \label{eq:Gamma}
  <\Gamma>_{op}(t) = \frac{1}{N^2} \left[1-2\left(1-\frac{\Delta
        O_c}{L}\right)^2\Theta\left(L-\Delta O_c\right) \right]\, ; 
\end{equation}
where $\Theta$ is the Heaviside function, $\Theta(x)=1$ if $x>0$ and zero
otherwise. To conclude we need to compute the time average of
$<\Gamma>_{op}(t)$. Using 
once again the \lq\lq triangle--like\rq\rq convergence assumption for the
opinions, i.e. $L(t)=1-t/T_c$, where $T_c$ is the opinion convergence time, we
get: 
\begin{equation}
  \label{eq:GammaTc}
  \gamma_{ef\!f}=\lim_{t\rightarrow \infty}\frac{1}{t} \int_0^t d\tau
  \frac{1}{N^2} \left[ 1-2\left(1-2\frac{\Delta
        O_cT_c}{T_c-\tau}+\left(\frac{\Delta
          O_cT_c}{T_c-\tau}\right)^2\right)\Theta\left(\frac{T_c-\tau}{T_c}-\Delta O_c\right)\right]\, ,  
\end{equation}
This integral can be explicitly solved to give:
\begin{equation}
  \label{eq:gammaefffinal}
  \gamma_{ef\!f}=
 \begin{cases}
 \frac{1}{N^2}\left(1+\frac{T_c\rho}{T_{\alpha}}\right) &  \text{if
     $T_\alpha>T_c$}\\ 
  \frac{\rho+1}{N^2} & \text{if  $T_\alpha<T_c$} \, ,
  \end{cases}
\end{equation}
where $\rho=-(1+2 \Delta O_c \log(\Delta O_c) - \Delta O_c^2 )$.

\end{document}